\documentclass[mathleft
]{an}
\usepackage{graphicx}
\usepackage{times}
\newcommand{\aaps}{A\&AS }

\newcommand{\jrasc}{JRASC}
\newcommand{\nat}{Nature}

\begin{document}

\Pagespan{789}{}
\Yearpublication{2006}%
\Yearsubmission{2005}%
\Month{11}%
\Volume{999}%
\Issue{88}%

\title{Photometric analysis of the eclipsing binary 2MASS\,19090585 +4911585\thanks{Based on observations obtained with telescopes of the University Observatory Jena, which is operated by the Astrophysical Institute of the Friedrich-Schiller-University Jena.}}

\author{St. Raetz\inst{1}\fnmsep\thanks{Corresponding author:
  \email{straetz@astro.uni-jena.de}\newline}
\and M. Va{\v n}ko\inst{1}
\and M. Mugrauer\inst{1}
\and T. O. B. Schmidt\inst{1}
\and T. Roell\inst{1}
\and T. Eisenbeiss\inst{1}
\and M. M. Hohle\inst{1,3}
\and \\ A. Koeltzsch\inst{1}
\and Ch. Ginski\inst{1}
\and C. Marka\inst{1}
\and M. Moualla\inst{1}
\and N. Tetzlaff\inst{1}
\and Ch. Broeg\inst{2}
\and R. Neuh{\"a}user\inst{1}
}
\titlerunning{Photometric analysis of 2MASS\,19090585+4911585}

\institute{
Astrophysikalisches Institut und Universit{\"a}ts-Sternwarte Jena, Schillerg{\"a}{\ss}chen 2-3, 07745 Jena, Germany
\and 
Space Research and Planetary Sciences, Physikalisches Institut, University of Bern, Sidlerstra{\ss}e 5, 3012 Bern, Switzerland
\and
Max Planck Institute for Extraterrestrial Physics, Giessenbachstra{\ss}e, 85748 Garching, Germany}

\received{2008 Dec 9}
\accepted{2009 Apr 11}
\publonline{2009 May 30}


\abstract{We report on observations of the eclipsing binary 2MASS\,19090585+4911585 with the 25\,cm auxiliary telescope of the University Observatory Jena. We show that a nearby brighter star (2MASS\,19090783+4912085) was previously misclassified as the eclipsing binary and find 2MASS\,19090585+4911585 to be the true source of variation. We present photometric analysis of $VRI$ light curves. The system is an overcontact binary of W UMa type with an orbital period of (0.288374\,$\pm$\,0.000010)\,d.}

\maketitle

\section{Introduction}

The discovery of the first close-in extrasolar planet around a sun-like star (Mayor \& Queloz 1995) revived the idea that transits of exoplanets in front of their parent stars could be observable (Struve 1952). From 1995 on several groups developed photometric search programs.\\ Since the discovery of the first transiting extrasolar planet (Charbonneau et al. 2000; Henry et al. 2000; Mazeh et al. 2000) over 20 ground-based projects and several space missions using the transit method have started. Most of them (e.g. TrES: Alonso et al. 2004; HAT: Bakos et al. 2004; SuperWASP: Christian et al. 2006; XO: McCullough et al. 2006) are designed as wide-field transit searches using small aperture telescopes with a diameter of less than 15\,cm.\\ The advantage of wide-field surveys is that they monitor several thousand stars in several hundred hours of observing. Besides planetary transit detections, wide-field photometric transit searches will also find all kinds of new and existing variable stars, including eclipsing binaries. For example, Devor et al. (2008) presented a catalog of 773 eclipsing binaries found in ten fields of the TrES survey.\\ In this paper, we describe the analysis of one eclipsing binary system located in the field of view around a star with a known transiting planet.  We could identify the type of eclipsing binary and estimate the physical properties, including the temperatures of the stellar components, as well as their mass ratio and orbital inclination.\\ The study of such objects helps to achieve a better understanding of how stars evolve and interact.

\section{Observations and Data Reduction}
\label{observations}

\begin{table}
\caption{Observation log of the field around TrES-2, where 2MASS\,19090585+4911585 was detected.}
\label{observation_log}
\begin{tabular}{lccc}\hline
Date & Interval [UT] & Filter & n$^{a}$ \\
\hline
2007 Mar 13 & 00:17:19\,-\,03:56:17 & $I$ & 145 \\
2007 Mar 27 & 21:11:46\,-\,00:36:06 & $R$ & 127 \\
2007 May 03 & 20:58:15\,-\,01:08:56 & $I$ & 134 \\
2007 May 24 & 20:49:24\,-\,22:01:26 & $I$ & 48 \\
2007 Jul 17 & 00:20:26\,-\,02:03:48 & $I$ & 70 \\
2007 Jul 17 & 21:25:49\,-\,22:39:01 & $I$ & 49 \\
2007 Jul 25 & 22:22:05\,-\,23:58:27 & $I$ & 59\\
2007 Aug 01 & 20:11:51\,-\,00:32:42 & $I$ & 160\\
2007 Aug 05 & 20:58:55\,-\,23:27:57 & $I$ & 96\\
2007 Aug 06 & 20:23:55\,-\,23:15:27 & $I$ & 110\\
2007 Aug 07 & 20:03:29\,-\,21:31:56 & $I$ & 64\\
2007 Aug 22 & 19:48:52\,-\,00:03:23 & $I$ & 149\\
2007 Aug 23 & 21:26:57\,-\,03:18:40 & $I$ & 217\\
2007 Aug 24 & 19:22:31\,-\,00:19:45 & $I$ & 188\\
2007 Aug 27 & 21:09:43\,-\,22:24:21 & $I$ & 42\\
2007 Aug 29 & 19:22:30\,-\,00:06:30 & $I$ & 68\\
2007 Sep 05 & 19:23:30\,-\,21:04:32 & $I$ & 62\\
2007 Sep 14 & 00:20:57\,-\,01:53:23 & $I$ & 60\\
2007 Sep 15 & 22:58:22\,-\,03:16:45 & $I$ & 159\\
2007 Sep 19 & 01:03:57\,-\,02:07:01 & $I$ & 41\\
2007 Sep 19 & 18:40:58\,-\,22:41:48 & $I$ & 145\\
2007 Sep 23 & 18:07:51\,-\,23:01:04 & $I$ & 189 \\
2007 Sep 25 & 18:13:09\,-\,22:51:46 & $I$ & 82 \\
2007 Oct 10 & 17:55:29\,-\,18:44:25 & $I$ & 33 \\
2007 Oct 13 & 22:15:23\,-\,02:18:51 & $I$ & 148 \\
2007 Oct 14 & 17:37:49\,-\,22:01:26 & $I$ & 167 \\
2007 Oct 16 & 17:31:53\,-\,19:17:39 & $I$ & 61 \\
2007 Oct 31 & 17:06:56\,-\,20:55:45 & $I$ & 132 \\
2008 Apr 23 & 21:50:49\,-\,02:35:01 & $V$ & 62 \\
& & $R$ & 61 \\
& & $I$ & 61\\
2008 May 03 & 22:20:03\,-\,01:45:59 &  $V$ & 40 \\
& & $R$ & 40 \\
& & $I$ & 40\\
2008 Jun 26 & 22:13:17\,-\,01:47:02 & $I$ & 145 \\
2008 Jul 28 & 23:04:01\,-\,01:26:18 & $V$ & 33 \\
& & $R$  & 32 \\
& & $I$ & 32\\
2008 Aug 04 & 23:40:24\,-\,02:28:01 & $V$ & 32 \\
& & $R$ & 30 \\
& & $I$ & 31\\
\hline
\end{tabular}
\\
$^{a}$ Number of images\\
\end{table}
During our observations of the transiting extrasolar planet host star TrES-2 at the University Observatory Jena (Raetz et al. 2009a, Raetz et al. 2009b), we found that 2MASS190 90585+4911585 exhibits some brightness variations with relatively large amplitude. \\ Our observations at the University Observatory Jena were performed with the 25\,cm Cassegrain auxiliary telescope \\ equipped with the optical CCD-camera CTK (for more details see Mugrauer 2009). \\ We started our observations in November 2006. For our TrES-2 observations, started in March 2007, we used 41 nights from March 2007 to August 2008. In 33 nights we could use these observations to analyze 2MASS\,19090585 +4911585 as well. All observations of the TrES-2 host star were carried out with Bessell $I$ filter with an exposure time of 60\,s for each frame. We achieve a mean cadence of the data points of 1.4\,min (readout time of the CTK around 24\,s).\\ In order to derive color information of 2MASS\,19090585 +4911585, besides our $I$-band observations we started $V$- and $R$-band photometry in April 2008. All observations are summarized in Table \ref{observation_log}. \\ 2MASS\,19090585+4911585 is a relatively faint star ($V$\,$\approx$ 15\,mag) and exposure times of our observations were selected mostly for the TrES-2 host star. The mean photometric accuracy of the observations in $V$ is $\approx$\,0.060\,mag, in $R \approx$\,0.045\,mag and in $I \approx$\,0.044\,mag.\\
\begin{figure*}
  \centering
  \includegraphics[width=0.9\textwidth]{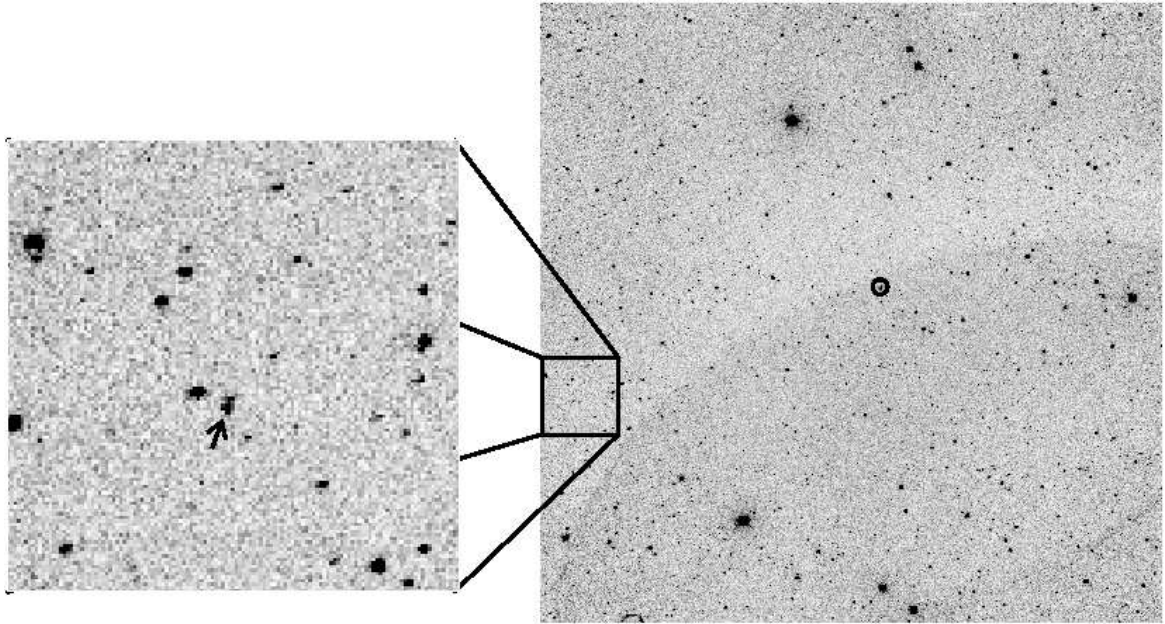}
  \caption{A zoom into the field of view of the CTK ($I$-band, exposure time 60\,s). The circle marks the TrES-2 host star, the arrow 2MASS\,19090585+4911585. North is up; east is to the left. The field size is 37.7$'\times$37.7$'$ on the right and 4.7$'\times$4.7$'$ on the left. The superimposed ringlike structure is caused by reflections of a nearby bright star.}
  \label{position_EB}
\end{figure*}
We calibrate the images of our target field using the standard IRAF\footnote{IRAF is distributed by the National Optical Astronomy Observatories, which are operated by the Association of Universities for Research in Astronomy, Inc., under cooperative agreement with the National Science Foundation.} procedures \textit{darkcombine}, \textit{flatcombine} and \textit{ccdproc}. We did not correct for bad pixel (see Raetz et al. 2009c).

\section{Photometry}

\begin{figure}
  \centering
  \includegraphics[width=0.4\textwidth]{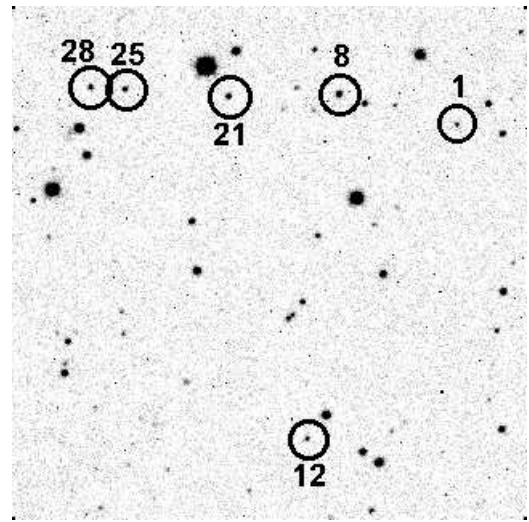}
  \caption{A 11.8$'\,\times$\,11.8$'$ CTK-120s-$I$-band image section of standard star field \#1 defined by Galad\'i-Enr\'iquez et al. (2000). The circles mark the used standard stars (brightness of the standard stars in the same range as for 2MASS\,19090585+4911585). For a definition of the labels see Galad\'i-Enr\'iquez et al. (2000)}.
  \label{standard_stars}
\end{figure}
\begin{figure}
  \centering
  \includegraphics[width=0.48\textwidth]{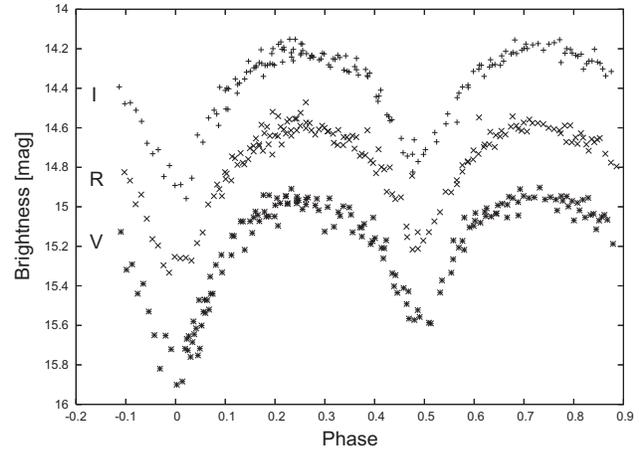}
  \caption{The $VRI$ light curves of 2MASS\,19090585+4911585 for the four nights in 2008 with quasi-simultaneous $VRI$-band observations.}
  \label{LC_mag}
\end{figure}
\begin{figure}
  \centering
  \includegraphics[width=0.48\textwidth]{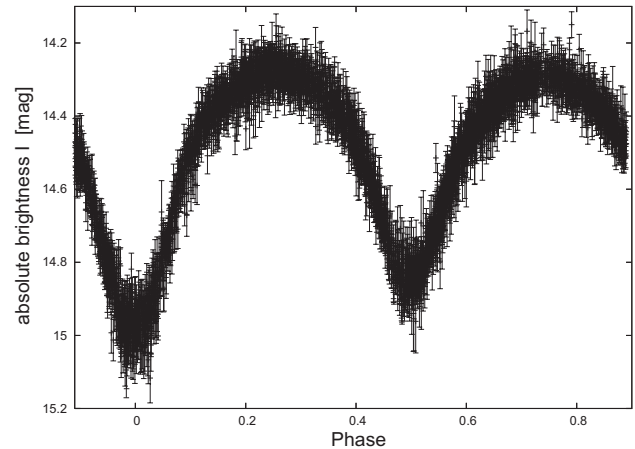}
  \caption{All our 3124 individual $I$-band observations of 2MASS\,19090585+4911585 obtained in 2007 and 2008 phased with the ephemeris given in eq. \ref{ephemeris}.}
  \label{LC_I_mag}
\end{figure}

\subsection{Differential photometry}

After calibrating all images, we perform aperture photometry on a maximum number of stars by using the IRAF task \textit{chphot} (see Raetz et al. 2009c). Because 2MASS\,19090585 +4911585 is a relatively faint star we did the photometry only on stars with similar brightness. We used an aperture of radius 2 pixels (4.4$''$) and an annulus for sky subtraction ranging in radius from 15 to 20 pixels, centered on each star. We choose this very large gap of 13 pixels to exclude the brighter star (2MASS\,19090783+4912085) next to 2MASS\,19090585+4911585 (see the left part in Fig. \ref{position_EB}) from the measurements. \\ For differential photometry we use an algorithm that calculates an artificial comparison star of all measured stars by taking the weighted average of them. To compute the best possible artificial comparison star we successively sort out all stars with low weights (stars that are not present on every image, stars with low signal-to-noise ratio (S/N) and variable stars which would introduce false signals to the data). As result, we get an artificial comparison star made of the most constant stars with the best S/N of the chosen sample of stars (Broeg et al. 2005, Raetz et al. 2009c). We performed this analysis individually for every night. To obtain comparable values for every night, we use for every sequence the same 23 comparison stars to calculate the artificial comparison star.\\ Finally the artificial comparison star is compared to 2MASS 19090585+4911585 to get the differential magnitudes for every image.

\subsection{Absolute photometry}

To derive absolute photometric $V$, $R$ and $I$ magnitudes we observed 2MASS\,19090585+4911585 on 2008 July 28, a photometric night. In order to perform the transformation to the standard system we observed standard star field \#1 defined by Galad\'i-Enr\'iquez et al. (2000) which is located around the two Landolt standard stars SA\,44\,28 and SA\,44 113 (Landolt 1983). Altogether we took four images of the standard star field at different airmasses and a sequence of 32 2MASS\,19090585+4911585 images in each filter. We measured the instrumental magnitudes of 6 secondary standard stars ($UBVRI$ photometry given in Galad\'i-Enr\'iquez et al. (2000), indentification chart shown in Fig. \ref{standard_stars}) in each frame and calculated the zero point correction $c$ and the first order extinction coefficient $k$. The results for each filter are shown in Table \ref{c_and_k}.\\ During the analysis it turned out that the light of the faint object near 2MASS\,19090585+4911585 (see Fig. \ref{position_EB}, left image) contaminates the PSF of the eclipsing binary. We chose a larger aperture for the absolute photometry than for differential photometry (5 pixels instead of 2 pixels) to include this faint star in our measurements. With the help of the flux ratio of the two stars we were able to calculate the individual brightnesses.\\ Using a sequence of 32 images we determined in which part of the light curve the measurement took place. With the help of the relative magnitudes we calculated the magnitudes in the two minima. We finally derived the $V$, $R$ and $I$ magnitudes for the primary and the secondary component of 2MASS\,19090585+4911585 on every image. Table \ref{abs_photometry_result} shows the resulting magnitudes for the primary (secondary minimum) and the secondary (primary minimum) component obtained from the average of all 32 individual measurements. The errors given in Table \ref{abs_photometry_result} correspond to the standard deviation of these measurements. We point out that the magnitudes in the two minima are only approximately the magnitudes of both components. Because the inclination is $<$\,90\degr\, the light of both components is slightly contaminated by the other component. We simultaneously determined the spectral types and measured the extinction to be $A_{\mathrm{V}}<0.1\,$mag. 

\section{Minima times and period determination}
\label{period}

\begin{table}
\centering
\caption{zero point correction $c$ (normalized to 1\,s exposure time) and first order extinction coefficient $k$ for each filter}
\label{c_and_k}
\begin{tabular}{cr@{\,$\pm$\,}lr@{\,$\pm$\,}l}
\hline
Filter & \multicolumn{2}{c}{$c$} & \multicolumn{2}{c}{$k$} \\ \hline 
$V$ & 18.96 & 0.03 & 0.27 & 0.02 \\
$R$ & 18.94 & 0.02 & 0.21 & 0.02 \\
$I$ & 18.42 & 0.02 & 0.16 & 0.01 \\ \hline
\end{tabular}
\end{table}
\begin{table*}
\caption{Derived $VRI$ magnitudes and colors for both components of 2MASS\,19090585+4911585}
\label{abs_photometry_result}
\begin{tabular}{cr@{\,$\pm$\,}lr@{\,$\pm$\,}lr@{\,$\pm$\,}lr@{\,$\pm$\,}lr@{\,$\pm$\,}lc}
\hline
component & \multicolumn{2}{c}{$V$} & \multicolumn{2}{c}{$R$} & \multicolumn{2}{c}{$I$} & \multicolumn{2}{c}{$V\,-\,R$} & \multicolumn{2}{c}{$V\,-\,I$} & spectral type $^{a}$ \\ \hline
primary & 15.76 & 0.06 & 15.29 & 0.04 & 14.92 & 0.08 & 0.47 & 0.07 & 0.84 & 0.10 & G3\,-\,K2 \\
secondary & 15.95 & 0.06 & 15.41 & 0.04 & 15.08 & 0.08 & 0.55 & 0.14 & 0.90 & 0.15 & G4\,-\,K4.5 \\
\hline
\end{tabular}
\begin{flushleft}
$^{a}$Spectral type derived from the color index based on Kenyon \& Hartmann (1995)
\end{flushleft}
\end{table*}
Our observations enable us to determine 23 minima times of 2MASS\,19090585+4911585 - 10 primaries and 13 secondaries. To determine the time of the minima we fit a gaussian curve on the observed minima. The best fit was derived by minimization of $\chi^{2}$. For nights with simultaneous $VRI$-band observations we determined the times of minima separately for all filters and computed the weighted average. We summarize the observed minima times in Table \ref{minima_times}. We give the 1-$\sigma$ error bars.\\ We used the listed minima times to determine the ephemeris using the linear function for the heliocentric Julian date of minimum (Equation \ref{Elements}) where epoch $E$ is an integer and $T_{0}$ the time of minimum at epoch 0.
\begin{equation}
\label{Elements}
T_{\mathrm{Min}}=T_{0}+P\cdot E
\end{equation}
For the determination of the ephemeris we set our first observed minimum time as $T_{0}$. We performed a fourier analysis of the data to derive a preliminary period. With this two values we could calculate the epoch of the observed minima.\\ To improve the ephemeris we plotted the minima times over the epoch and did a linear $\chi^{2}$-fit. We got the best $\chi^{2}$ with a time of minimum to epoch 0 of $T_{0}\,=\,(2454172.643106\,\pm\,0.000080)\,\mathrm{d}$ and an orbital period of $P\,=\,(0.288374\,\pm\,0.000010)\,\mathrm{d}$. \\ The ephemeris which represents our measurements best is (according to the primary Minima)
\begin{equation}
\label{ephemeris}
T_{\mathrm{MinI}}\,=\,(2454172.643106\,+\,E\cdot 0.288374)\,\mathrm{d}
\end{equation}
The resulting $VRI$ light curves of 2MASS\,19090585+4911 585 phased with the ephemeris given in eq. \ref{ephemeris} are shown in Fig. \ref{LC_mag}, all $I$-band observations obtained in 2007 and 2008 in Fig. \ref{LC_I_mag}.

\section{Light curve analysis}
\label{analysis}

\begin{figure*}
  \centering
  \includegraphics[width=0.9\textwidth]{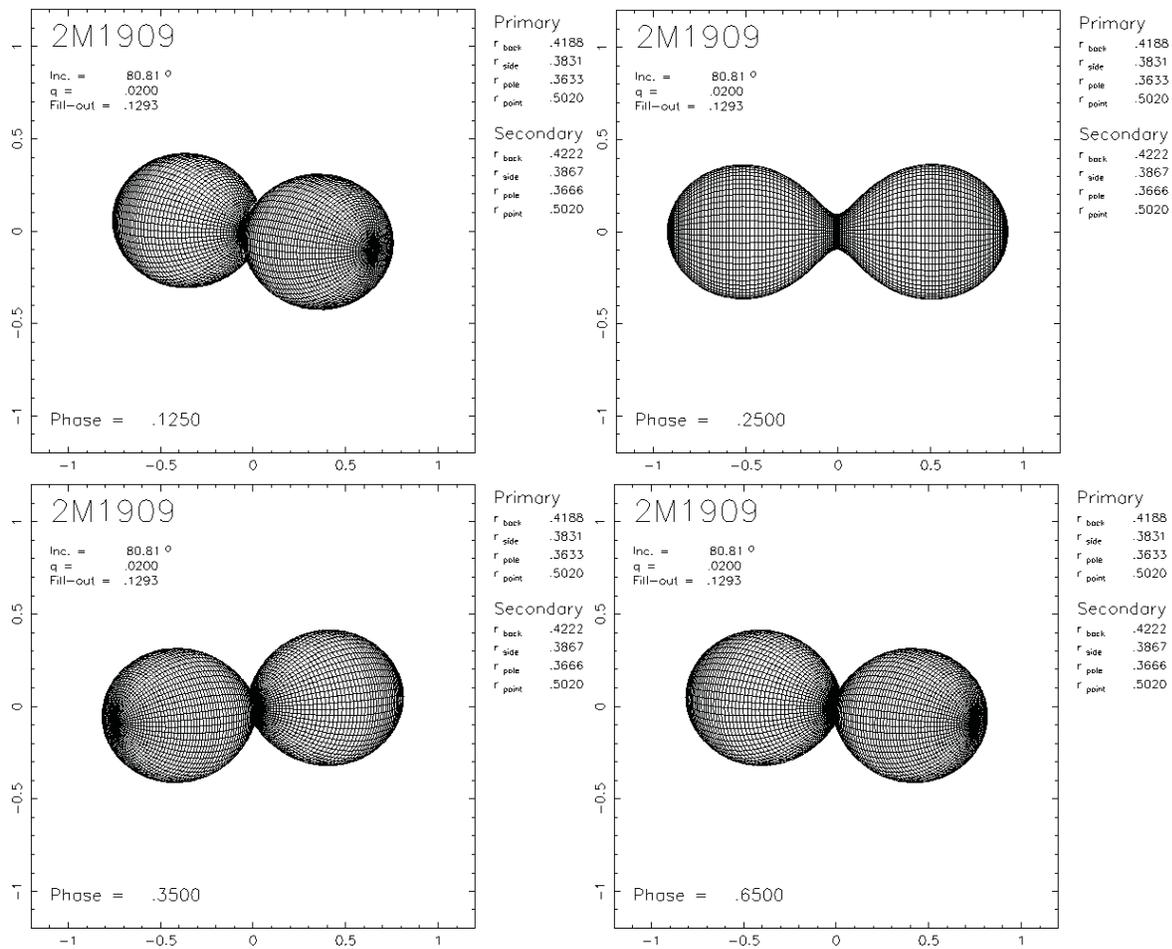}
  \caption{3D model of 2MASS\,19090585+4911585 in four different orbital phases.}
  \label{3D_model}
\end{figure*}

Our observations show symmetric light curves (hereafter LCs) in all passbands with brightnesses in both maxima at the same level. For our analysis we used only the LC of 2008 April 23, 2008 May 3, 2008 July 28 and 2008 August 4 because of quasi-simultaneous observations in all three passbands. \\ The analysis of LCs was performed by the software package PHOEBE (Pr{\v s}a and Zwitter, 2005). It is a modeling package for eclipsing binary stars which is based on the widely used WD (Wilson and Deviney) program (Wilson and Deviney, 1971; Wilson, 1979; 1990).\\ In the process of LC analysis of the binary we have adopted the following parameters: the mean temperature of the primary component, $T_{\mathrm{1}}$\,=\,5800\,K based on the ($V-R$) color index which corresponds within the error bars to G4 spectral type (Kenyon \& Hartmann, 1995), the coefficients of gravity darkening $g_{\mathrm{1}}\,=\,g_{\mathrm{2}}$\,=\,0.32 (Lucy, 1967) and the bolometric albedo coefficients of $A_{\mathrm{1}} = A_{\mathrm{2}}$\,=\,0.5 (Rucinski, 1969). These are appropriate values for the convective envelopes. We assumed the same values for gravity darkening and bolometric albedo for the primary and the secondary component because both have similar brightnesses and spectral types. A Kurucz (1993) model for stellar atmospheres was applied to the stars assuming solar composition. The second order bolometric and monochromatic limb darkening coefficients for logarithmic law were interpolated from van Hamme (1993) tables.\\ The weights of individual data points were established as 1/$\sigma^{2}$, where $\sigma$ is the standard error of the point derived during the photometric measurement.\\ We performed a fit of all three ($VRI$) LCs simultaneously. Based on the shape of the LCs, we set the PHOEBE solution to overcontact binary which is not in thermal contact. The initial values of optimized (free) parameters were determined manually to give a visually good fit with our data. Using these values a correct solution was quickly found. We have run differential corrections fit until output corrections were smaller than the errors of the fitted parameters. \\ According to Binnendijk (1957), dividing of W UMa systems into A and W-subtypes, we concluded that this system is most likely a W-subtype because of G4 spectral type and short orbital period. This means that the primary minimum is caused by occultation of the less massive and hotter primary component by the more massive but cooler secondary one. Because of the W-type definition, the mass ratio is $q>1$.
\begin{table}
\centering
\caption{Times of the primary (I) and secondary (II) minima of 2MASS\,19090585+4911585}
\label{minima_times}
\begin{tabular}{lr@{\,$\pm$\,}ll}
\hline
Date & \multicolumn{2}{c}{Minimum time [HJD]} & Type \\ \hline
2007 Mar 13 & 2454172.6427 & 0.0005 & I \\
2007 Mar 27 & 2454187.4952 & 0.0007 & II \\
2007 May 03 & 2454224.4075 & 0.0005 & II \\
2007 Jul 25 & 2454307.4582 & 0.0004 & II \\
2007 Aug 01 & 2454314.3788 & 0.0004 & II \\
2007 Aug 05 & 2454318.4149 & 0.0004 & II \\
2007 Aug 06 & 2454319.4260 & 0.0003 & I \\
2007 Aug 22 & 2454335.4310 & 0.0003 & II \\
2007 Aug 23 & 2454336.4395 & 0.0003 & I \\
2007 Aug 23 & 2454336.5837 & 0.0003 & II \\
2007 Aug 24 & 2454337.4485 & 0.0003 & II \\
2007 Aug 29 & 2454342.4953 & 0.0005 & I \\
2007 Sep 16 & 2454359.5093 & 0.0003 & I \\
2007 Sep 19 & 2454363.4026 & 0.0003 & II \\
2007 Sep 23 & 2454367.2946 & 0.0004 & I \\
2007 Sep 23 & 2454367.4396 & 0.0005 & II \\
2007 Oct 13 & 2454387.4810 & 0.0004 & I \\
2007 Oct 14 & 2454388.3466 & 0.0004 & I \\
2007 Oct 31 & 2454405.2192 & 0.0005 & II \\
2007 Oct 31 & 2454405.3606 & 0.0005 & I \\
2008 Apr 24 & 2454580.5474 & 0.0006 & II \\
2008 May 03 & 2454590.4976 & 0.0004 & I \\
2008 Jun 27 & 2454644.5662 & 0.0005 & II \\
\hline
\end{tabular}
\end{table}
\begin{table}
\centering
\caption{Photometric elements of 2MASS\,19090585+4911585 and their standard errors: i\,-\,orbital inclination, q\,-\,photometric mass ratio, $\Omega$\,-\,surface potential, T$_{1}$\,-\,temperature of the primary component, T$_{2}$\,-\,temperature of the secondary component, r$_{1}$,r$_{2}$\,-\,volume mean fractional radii, $L_{\mathrm{1}}$/\,($L_{\mathrm{1}}$+$L_{\mathrm{2}}$)$_{\mathrm{V,R,I}}$\,-\,relative luminosities, $\sum$\,w(O-C)$^{2}$\,-\,normalized weighted sum of squares of residuals for all LCs.}
\label{photometric_elements}
\begin{tabular}{lr@{\,$\pm$\,}l}
\hline
Parameter & \multicolumn{2}{c}{Value} \\ \hline
$i$ [\degr] & 79.13 & 0.43 \\
$q$ [\degr] & 1.079 & 0.008 \\
$\Omega_{\mathrm{1}}$\,=\,$\Omega_{2}$ & 3.792 & 0.019 \\
$T_{\mathrm{1}}$ [K] & \multicolumn{2}{c}{5800$^{a}$}  \\
$T_{\mathrm{2}}$ [K] & 4683 & 51 \\
$r_{\mathrm{1}}$ & 0.316 & 0.010 \\
$r_{\mathrm{2}}$ & 0.320 & 0.013 \\
$L_{\mathrm{1}}$/\,($L_{\mathrm{1}}$+$L_{\mathrm{2}}$)$_{\mathrm{V}}$ & 0.595 & 0.053 \\
$L_{\mathrm{1}}$/\,($L_{\mathrm{1}}$+$L_{\mathrm{2}}$)$_{\mathrm{R}}$ & 0.529 & 0.047 \\
$L_{\mathrm{1}}$/\,($L_{\mathrm{1}}$+$L_{\mathrm{2}}$)$_{\mathrm{I}}$ & 0.498 & 0.051 \\
$\sum\,w(O-C)^{2}$ & \multicolumn{2}{c}{1.983} \\
\hline
\end{tabular}
\\
$^{a}$ fixed according to spectral type (Kenyon \& Hartmann, 1995)
\end{table}
\begin{figure} 
 \centering
  \includegraphics[width=0.48\textwidth]{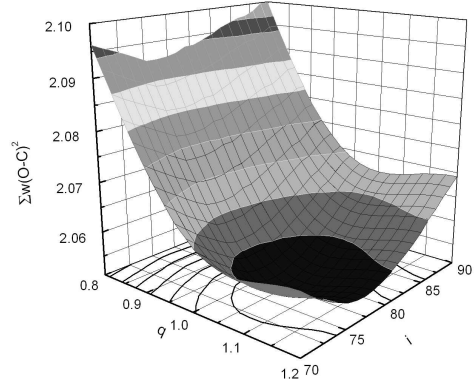}
  \caption{The variations of the normalized weighted sum of squares of residuals $\sum$\,w(O-C)$^{2}$ for the light curves for different values of mass ratio $q$ and orbital inclination $i$.}
  \label{sum_w(O-C)^2}
\end{figure}
\begin{figure}
  \centering
  \includegraphics[width=0.48\textwidth]{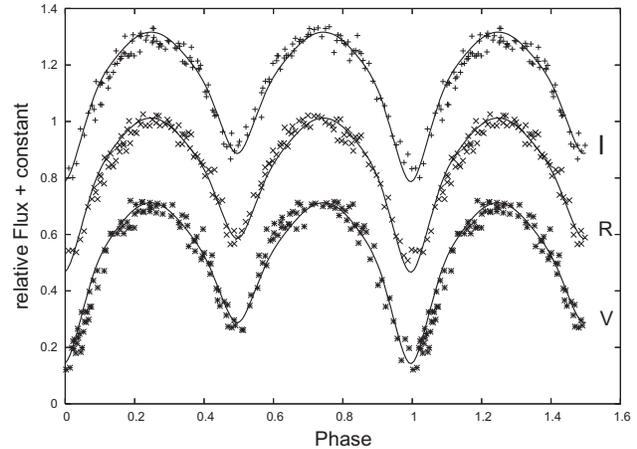}
  \caption{The best fits of $VRI$ light curves of 2MASS\,19090585+4911585 for the data which were used for the analysis.}
  \label{LC_fits}
\end{figure}
\\ To find correct values of mass ratio $q$ and orbital inclination $i$, we computed the normalized weighted sum of squares of residuals $\sum$\,w(O-C)$^{2}$ for fixed values of $q$ in the interval 0.8\,-\,1.2 and $i$ in the interval 60\degr\,-\,90\degr and looked for a global minimum. A detailed view of the variations of the sum of squares around the global minimum is shown in Fig. \ref{sum_w(O-C)^2}. While fitting the LCs in all passbands we obtained photometric parameters for the system which are listed in the Table \ref{photometric_elements}. Because of the range of possible spectral types these results should be treated carefully. The best fit of LCs for all passbands are depicted in Fig. \ref{LC_fits}. A corresponding 3D model of the system is shown in Fig. \ref{3D_model}.\\ The amplitude of light variations of 2MASS\,19090585+491 1585 is $\sim$\,0.898\,mag in $V$, $\sim$\,0.841\,mag in $R$ and $\sim$\,0.799 mag in $I$ for the primary Minimum and $\sim$\,0.591\,mag in $V$, $\sim$\,0.595\,mag in $R$ and $\sim$\,0.598\,mag in $I$ for the secondary minimum.

\section{Erratum of a previous publication}

2MASS\,19090585+4911585 is located in the field of view of the Trans-atlantic Exoplanet Survey (TrES; Alonso et al. 2004). Devor et al. (2008) describe the identification and analysis of 773 eclipsing binaries found in data which were observed with Sleuth, one telescope of the TrES-network (O'Donovan et al. 2004). Sleuth is a 10\,cm telescope that images a 6\degr\,$\times$\,6\degr\, field of view onto a 2048\,$\times$ 2048 CCD camera (pixel scale $\sim$\,10.5$''$/Pixel).\\ To exclude that 2MASS\,19090585+4911585 was already discovered by the TrES-network we compared the photometric parameters we derived with the 773 eclipsing binaries published by Devor et al. (2008). We found an object whose orbital period is within the error bars identical to the one we give in section \ref{period}. Through comparison of the coordinates of 2MASS\,19090585+4911585 and the object with the same orbital period it turned out that Devor et al. (2008) refers to the brighter object on the left of 2MASS\,19090585+4911585 (see Fig. \ref{position_EB}) -- 2MASS\,190907 83+4912085. To check if they mean the same eclipsing binary but referring to the wrong star, we examined the amplitude of brightness variation. Our values which are given in section \ref{analysis} differ significantly from the estimates they gave. We conclude that the large pixel scale of Sleuth is not sufficient to resolve the true source of variation. On the CCD-camera of Sleuth 2MASS\,19090585+4911585 and the two nearby objects (Fig. \ref{position_EB}, left image) are convolved in one PSF as their relative distance corresponds to $<$\,2 pixel. We calculated the amplitude of flux variation with respect to the total flux of all three objects, converted this value into magnitudes and find a similar result as given by Devor et al. (2008).

\section{Summary and Conclusions}

Using the 25\,cm auxiliary Cassegrain telescope of the University Observatory Jena together with its optical CCD camera CTK, we observed the eclipsing binary 2MASS\,19090 585+4911585 in the field of view around the host star of the known transiting planet TrES-2. We found that instead of 2MASS\,19090783+4912085 which was previously thought to be the eclipsing binary 2MASS\,19090585+4911585 is the true source of variation.\\ From our observations in one photometric night we were able to derive absolute photometric $V$, $R$ and $I$ magnitudes. Hence, we determined the colors and the spectral types of both components.\\ Through analysis of altogether 23 minima we found the ephemeris to be $T_{\mathrm{Min}}$\,=\,(2454172.643106\,+\,E\, $\cdot$\,0.288374)\,d. \\ Our light curve analysis of $VRI$ photometric observations of 2MASS\,19090585+4911585 showed, that this system is an overcontact binary of W\,UMa type. Considering its spectral type and orbital period, we concluded that the primary minimum is an occultation (primary component is eclipsed by secondary one). Hence, this system belongs to an W-type of W\,UMa binaries. The amplitude of light variations is $\sim$\,0.90\,mag in $V$, $\sim$\,0.84\,mag in $R$ and $\sim$\,0.80\,mag in $I$. While assuming parameters like the temperature for the primary component, gravity darkening and bolometric albedo for both components and fitting the LCs in all passbands, we obtained photometric parameters for the system. The temperature difference between the components and the photometric mass ratio of both components is about 1100\,K and 1.08, respectively.

\acknowledgements
The authors would like to thank the GSH Observer Team for the nightly observations. SR and MV acknowledge support from the EU in the FP6 MC ToK project MTKD-CT-2006-042514. RN acknowledges general support from the German National Science Foundation (Deutsche Forschungsgemeinschaft, DFG) in grants NE 515/13-1, 13-2, and 23-1. AK acknowledges support from DFG in grant KR 2164/8-1. TR would like to thank the DFG for financial support (grant NE 515/23-1). TOBS acknowledges support from Evangelisches Studienwerk e.V. Villigst. TE and MMH acknowledge support from the DFG in SFB-TR 7 Gravitational Wave Astronomy. M.Moualla thanks the government of Syria for financial support. Furthermore, we thank the technical staff of the University Observatory Jena in Gro{\ss}schwabhausen, especially Tobias B{\"o}hm.

\end{document}